\documentclass[%
 aip, apl
 amsmath,amssymb,
reprint,%
]{revtex4-1}

\usepackage{graphicx}
\usepackage{dcolumn}
\usepackage{bm}

\usepackage{verbatim}

\usepackage[utf8]{inputenc}
\usepackage[T1]{fontenc}
\usepackage{mathptmx}

\usepackage{natbib}
\usepackage{xcolor}

\begin{document}
\preprint{AIP/123-QED}

\title[Edge mode engineering for optimal ultracoherent silicon nitride membranes]{Edge mode engineering for optimal ultracoherent silicon nitride membranes}

\author{E. Ivanov}
 \affiliation{%
Laboratoire Kastler Brossel, Sorbonne Universit\'e,  CNRS, ENS-Universit\'e PSL, Coll\`ege de France, 75005 Paris, France}
\author{T. Capelle}%
\affiliation{%
Laboratoire Kastler Brossel, Sorbonne Universit\'e,  CNRS, ENS-Universit\'e PSL, Coll\`ege de France, 75005 Paris, France}
\author{M. Rosticher}
\affiliation{D\'epartement de Physique, ENS-Universit\'e PSL, CNRS, 24 rue Lhomond, F-75005 Paris, France}
\author{J. Palomo}
\affiliation{D\'epartement de Physique, ENS-Universit\'e PSL, CNRS, 24 rue Lhomond, F-75005 Paris, France}\author{T. Briant}
\author{P.-F. Cohadon}
\author{A. Heidmann}
\author{T. Jacqmin}
\author{S. Del\'eglise}
\affiliation{%
Laboratoire Kastler Brossel, Sorbonne Universit\'e,  CNRS, ENS-Universit\'e PSL, Coll\`ege de France, 75005 Paris, France}

\date{\today}

\begin{abstract}
Due to their high force sensitivity, mechanical resonators combining low mechanical dissipation with a small motional mass are highly demanded in fields as diverse as resonant force microscopy, mass sensing, or cavity optomechanics.
``Soft-clamping'' is a phononic engineering technique by which mechanical modes of highly-stressed membranes or strings are localized away from lossy regions, thereby enabling ultrahigh-Q for ng-scale devices.
Here, we report on parasitic modes arising from the finite size of the structure which can significantly degrade the performance of these membranes.
Through interferometric measurements and finite-element simulations, we show that these parasitic modes can hybridize with the localized modes of our structures, reducing the quality factors by up to one order of magnitude.
To circumvent this problem, we engineer the spectral profile of the parasitic modes in order to avoid their overlap with the high-Q defect mode. 
We verify via a statistical analysis that this modal engineering reproducibly yields higher quality factors in fabricated devices, consistent with theoretically predicted values.
\end{abstract}

\maketitle




Mechanical resonators have been widely used in force sensing owing to the exceptional sensitivities that they can achieve in state-of-the-art devices \cite{Rugar2004,Hanay2012}.
In addition, the long coherence times of motional states of modern mechanical resonators, reaching the milliseconds \cite{Rossi2018}, can be exploited in quantum information to produce quantum memories\cite{Chu2018}, to couple otherwise incompatible quantum systems such as for optical-to-microwave photon conversion \cite{Andrews2014,Vainsencher2016,Jiang2020,Forsch2020}, or to study non-classical states of motion \cite{Marinkovic2018,Viennot2018,Shomroni2019a}.
Tensions silicon nitride (SiN) membranes are among the most popular platforms because of their ultrahigh quality factor $Q$.
This is enabled by a mechanism known as dissipation dilution: although the intrinsic quality factor $Q_\mathsf{intr}$ of the amorphous SiN is low (of the order of a few thousands\cite{Villanueva2014}), the ability to grow this material with high tensile stress can be exploited to obtain $Q \gg Q_\mathsf{intr}$\cite{Unterreithmeier2010} in high-aspect ratio structures.

In highly stressed membranes and strings, most of the elastic energy of vibrational modes is stored in lossless elongation rather than in bending, which causes dissipation in the presence of intrinsic material loss\cite{Zener1938,Yu2012}.
For a thin quasi-2D resonator vibrating out-of-plane with a profile $u(x, y)$, the energy lost per oscillation cycle $\Delta U$ is given by\cite{Yu2012} 
\begin{equation}\label{eqn:delta_u_def}
\Delta U = \frac{2\pi}{Q_\mathsf{intr}} U_\mathsf{bend}.
\end{equation}
Here, $U_\mathsf{bend}$ is the energy stored in bending, and is related to the mean curvature of the membrane $c(x, y) = \nabla^2 u(x,y)$ by
\begin{equation}\label{eqn:ubend}
U_\mathsf{bend} = \frac{Eh^3}{24(1-\nu^2)}\int 
c^2(x, y)
\hspace{1mm}dxdy,
\end{equation}
where $E$ is Young's modulus, $\nu$ is the Poisson ratio, and $h$ is the resonator thickness. 
On the other hand, the energy stored in elongation $U_\mathsf{elong}$ is related to the gradient of the mode profile $\mathbf{g}(x,y)=\mathbf{\nabla} u(x,y)$, and reads
\begin{equation}\label{eqn:uelong}
U_\mathsf{elong} = \frac{\sigma h}{2} 
\int
\mathbf{g}^2(x, y)
\hspace{1mm}dxdy,
\end{equation}
where $\sigma$ is the tensile stress.
In highly stressed systems where $U_\mathsf{elong} \gg U_\mathsf{bend}$, the effective gain in $Q$ is $Q/Q_\mathsf{intr} = U_\mathsf{elong} / U_\mathsf{bend}$.

In a model where only $u$ is subject to fixed boundary conditions at the borders of the resonators ($u|_\mathsf{borders}=0$), the mode shape is purely sinusoidal, and $u$ and its derivatives are independent of $\sigma$.
In that case, from Eqs. (\ref{eqn:ubend}) and (\ref{eqn:uelong}), the gain would be 
\begin{equation}\label{eqn:q_qintr}
\frac{Q}{Q_\mathsf{intr}} \propto \Lambda, 
\hspace{2mm} \mathsf{with} \hspace{2mm} 
\Lambda = \frac{\sigma l^2}{Eh^2},
\end{equation}
where $l$ is the typical lateral dimension of the mode.
However, the hard clamping boundary conditions in thin-film resonators imposes an additional constraint of $\mathbf{n}\cdot\mathbf{g}|_\mathsf{borders}=0$, where $\mathbf{n}$ is a vector normal to the borders.
The mode profile thus acquires a dependence on $\sigma$, which we denote with an additional index in $u_\sigma$, $\mathbf{g}_\sigma$, and $e_\sigma$.
In particular, although $u_\sigma(x,y)$ and $\mathbf{g}_\sigma(x,y)$ uniformly converge towards the limiting sinusoidal profiles as $\sigma \rightarrow \infty$, $c_\sigma(x,y)$ exhibits a divergence close to the anchor points\cite{Yu2012}. 
The net result is that the bending energy varies as $U_\mathsf{elong}\propto \sqrt{\sigma}$ \cite{Unterreithmeier2009}, and a modified scaling for the quality factor $Q_\mathsf{hc}$ of such ``hard clamped'' systems is found\cite{Yu2012}:
\begin{equation}\label{eqn:qhc_qintr}
\frac{Q_\mathsf{hc}}{Q_\mathsf{intr}} \propto \sqrt{\Lambda}.
\end{equation}

In practice, SiN can be deposited with $\sigma$ up to 1.4 GPa.
This value can be pushed close to the ultimate yield limit of approximately 6 GPa by stress engineering \cite{Ghadimi2017a}. 
For Young's modulus of $E=270$ GPa and an aspect ratio $l/h\approx$ 1 mm $/$ 20 nm (representative of state-of-the-art devices\cite{Rossi2018}), $Q_\mathsf{hc}/Q_\mathsf{intr}\sim 10^3$, rather than the potential $10^6$ for the ideal case described by Eq. (\ref{eqn:q_qintr}).

To overcome this shortcoming, a method was recently developed which uses phononic engineering to make anchor losses negligible\cite{Tsaturyan2017}. 
By structuring the resonators with a phononic crystal (PnC), the modes become localized within a central defect region, such that the PnC exponentially attenuates the modes and the asymptotic divergence of $c_\sigma$ is suppressed. 
These so-called ``soft clamped'' resonators benefit from the full dissipation dilution described by Eq. (\ref{eqn:q_qintr}), resulting in $Q$ exceeding $10^8$ at room temperature\cite{Tsaturyan2017,Ghadimi2017}.

A critical consideration in the design of such resonators is the impact of finite size effects of the PnC.
Indeed, \textit{edge modes} (EMs) which are localized near the borders of the resonator may be supported.
These EMs must satisfy the hard clamping boundary condition, and therefore follow the suboptimal gain in $Q$ of Eq. (\ref{eqn:qhc_qintr}).
Crucially, if some frequencies of EMs are degenerate to those of the central \textit{defect modes} (DMs), they can form hybridized modes which are no longer soft clamped, and the ideal dissipation dilution is lost.
Since fabrication-related errors result in fluctuations of the mode frequencies, the onset of mode hybridization is unpredictable, leading to resonators with a large sample-to-sample variability in $Q$.

In this Letter, we study parasitic EMs in membranes patterned with a PnC.
We experimentally show that a dense spectrum of EMs arises in PnC membranes, and can that it can lie in the vicinity of DM eigenfrequencies. 
We verify this result with finite-element simulations, and show that the DM and the EMs can indeed become hybridized.
To overcome this, we propose an alternative design in which the frequencies of the parasitic modes are shifted away from the DM of interest.
Through a statistical study of both designs, we demonstrate that the average performance of PnC membranes can be improved by implementing edge mode engineering techniques, resulting in higher reproducibility in the DM $Q$.

\begin{figure*}[htbp]
\includegraphics[width=17cm]{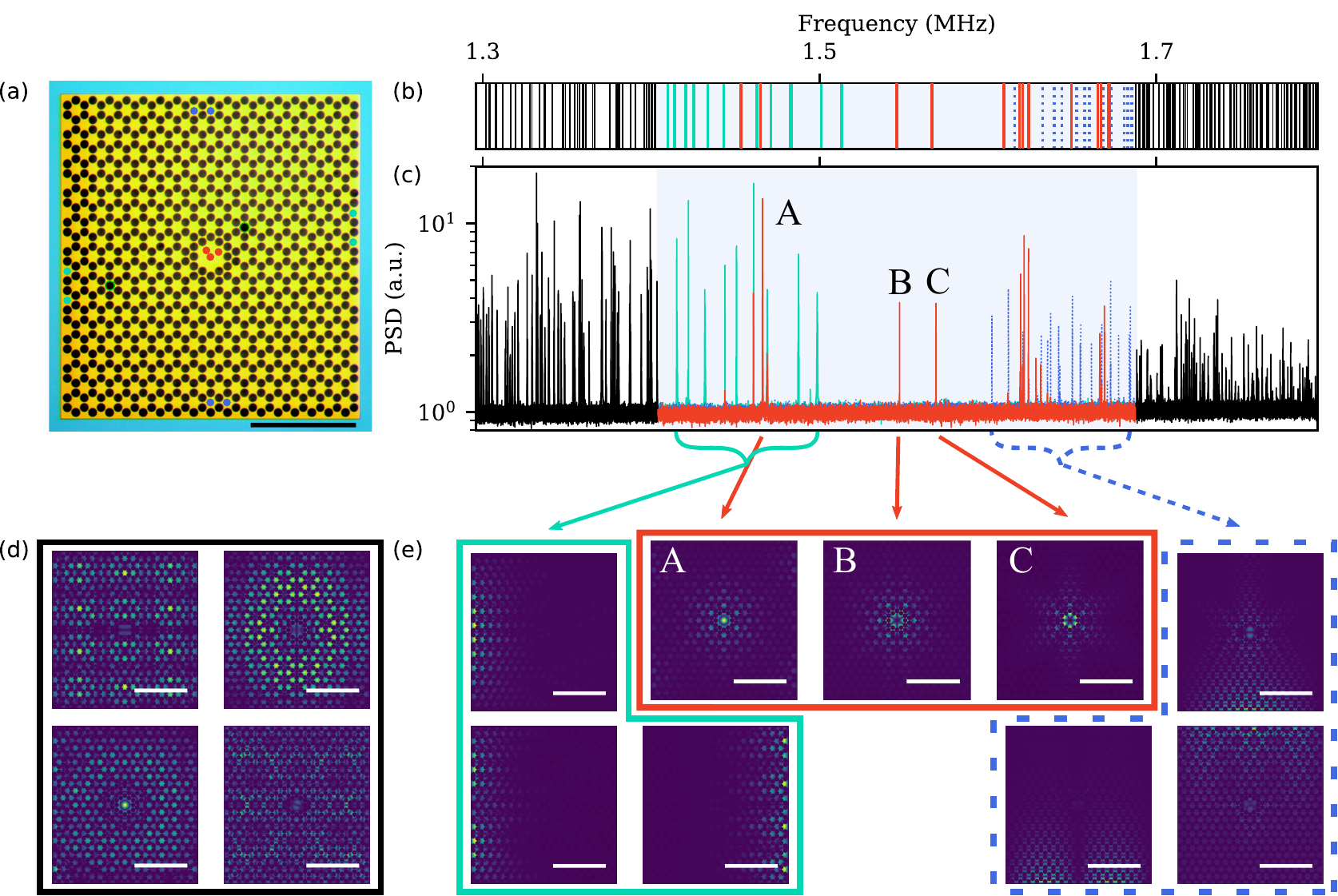}
\caption{\label{fig:gen1_memb}(a) Optical micrograph of a first generation PnC membrane. In yellow, the SiN is suspended above air, and in blue, the SiN lies above Si. 
The dots represent the points where the spectra shown in (c) are measured.
(b) Simulated spectrum near the PnC bandgap. 
(c) PSD spectrum, consisting overall of 4 spectra, measured at the vertical borders (cyan), the horizontal borders (blue), the defect (red), and within the PnC (black).
For both spectra, the bandgap is highlighted in light blue.
Absolute value of the simulated displacement fields of (d) four membrane modes and (e)
of selected examples of VEMs (left), DMs (center), and HEMs (right). 
The amplitude of the mode is arbitrary, with dark blue representing zero displacement, and yellow representing maximal displacement. All scale bars in this figure represent 1 mm.}
\end{figure*}

The samples presented in this work are fabricated from 4-inch Si (100) wafers purchased from Si-Mat\cite{simat}, on both sides of which a layer of SiN is deposited with a thickness of approximately 100 nm. 
Optical lithography and reactive ion etching are used on one side to form a hexagonal array of holes\cite{Tsaturyan2017}. 
A hot potassium hydroxide bath is then used to chemically back-etch the silicon substrate, releasing the PnC membrane; the details of the fabrication process are elsewhere\cite{Ivanov2020}.
Owing to the cumulative fabrication imprecision of optical lithography and pattern alignment, there is an overall imprecision in the membrane side length and PnC centering amounting to approximately $5$ $\mu$m.
An example of a ``first generation'' PnC membrane is shown in Fig. \ref{fig:gen1_memb}a, with dimensions of approximately 2.84 mm x 3.08 mm. 
The PnC holes have a radius of 42 $\mu$m, and the lattice parameter is 160 $\mu$m, creating a spectral bandgap for out-of-plane motion in the range of approximately 1.4 to 1.6 MHz.

Characterization is performed by spatially-resolved interferometry.
The thermal fluctuations of the membrane are probed with a 5 mW Nd:YAG laser, with a beam waist at the sample of approximately 10 $\mu$m. 
The power spectral densities (PSD) are measured with a balanced homodyne detection, effectively rejecting the technical noise of the laser below the shot-noise limit. 
The spectra are taken with a resolution bandwidth of $30$ Hz, which allows for a single-shot acquisition of the entire spectrum of interest.
The interferometer is set on a translating stage (Newport M-426 with LTA-HL actuator), such that the PSD can be recorded as a function of position, from which the displacement profiles of the eigenmodes can be reconstructed \cite{Barg2018}. 
All measurements are done at a pressure of a few nanobar, to ensure that gas damping remains negligible. 

\begin{figure*}[bt]
\begin{centering}
\includegraphics[width=17cm]{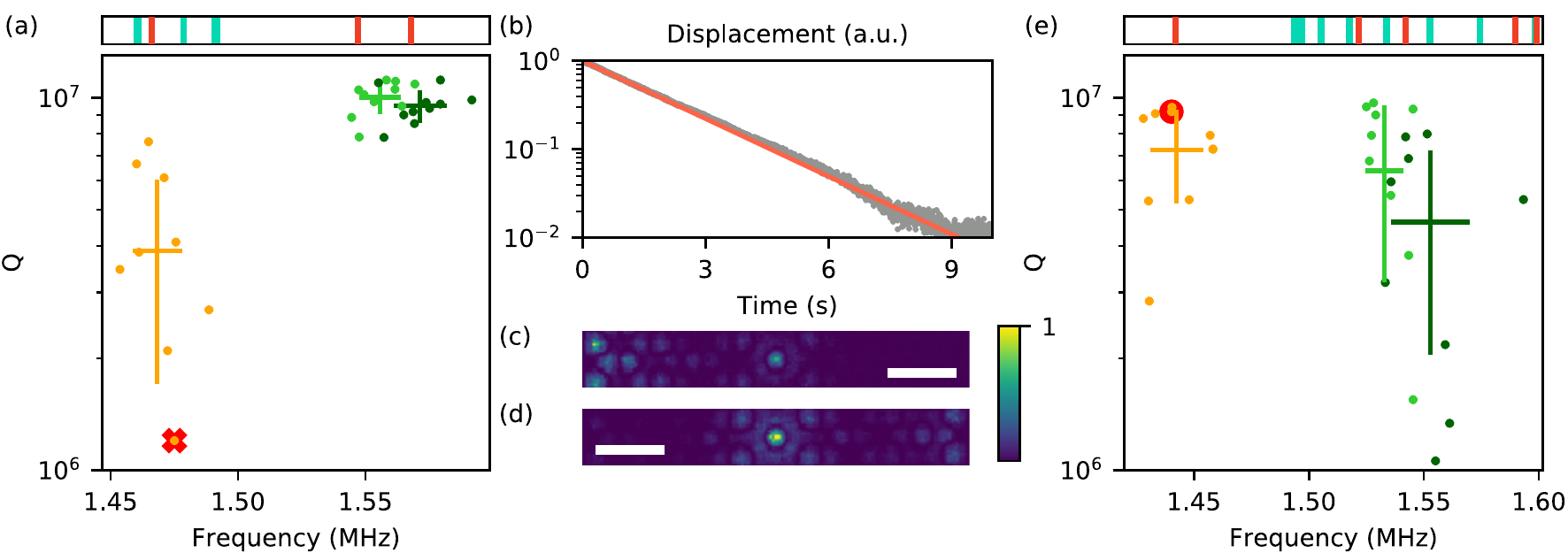}
\caption{\label{fig:qs_database} (a) Measured $Q$ values for modes A (yellow), B (light green), and C (dark green) for the first generation design. 
The simulated spectrum is reminded above. 
The red cross pinpoints the mode imaged in profiles (c) and (d).
(b) Example of ringdown measurement, from which $Q$ can be determined with a linear fit (orange).
(c)-(d) Mode profile images with significant displacement near the borders and the center. The displacement is plotted in arbitrary units, and the scale bar represents 1 mm. 
(e) Collected values of $Q$ for the second generation design, with the same color coding as (a). The simulated spectrum is reminded above. The red dot pinpoints the mode used for the ringdown measurement in (b).
In (a) and (d), the errors bars indicate the mean values of the quality factors of their corresponding color, and their standard deviation.}
\end{centering}
\end{figure*}

The quality factors $Q$ of our samples are measured by ringdown measurement. 
The membrane is first driven by modulating the intensity of the measurement beam.
After a few seconds, the intensity modulation is abruptly switched off; the interferometric signal is then used to continuously monitor the free evolution of the mode amplitude as a function of time. 
Finally, the envelope of the amplitude is fitted with a decaying exponential, from which $Q$ is deduced.
An example of a typical ringdown curve is shown in Fig. \ref{fig:qs_database}b.
We have checked that the measured $Q$ values are unaffected by the probe laser itself.

The simulated spectrum of a first-generation membrane around the bandgap of the PnC is shown in Fig. \ref{fig:gen1_memb}b.
The membranes are modelled by a quasi-3D system, using the finite-element solver COMSOL Multiphysics.
To faithfully render the richness of the spectrum in the presence of micro-imprecisions of fabrication, the PnC structure is shifted from the center of the membrane by 5 $\mu$m in each direction.
Outside of the highlighted bandgap, the modes are delocalized over the entire membrane; examples of such \textit{membrane modes} are shown in Fig. \ref{fig:gen1_memb}d.
Within the bandgap, we observe three different kinds of modes, categorized by their position on the membrane: DMs are localized near the central defect, while vertical EMs (VEMs) and horizontal EMs (HEM) are localized along the vertical and horizontal borders of the membrane, respectively.
Some illustrative examples of these kinds of modes are shown in Fig. \ref{fig:gen1_memb}e.
For the purposes of this work, only DMs labeled A, B, and C are considered.

Measuring a similar spectrum cannot be done in a single shot, as there is no single point at which all modes present significant displacement.
Several measurements are thus taken, at 13 different points on the membrane, pinpointed in Fig. \ref{fig:gen1_memb}a.
They are grouped into 4 categories, based on their location: on the defect, along the horizontal borders, along the vertical borders, and within the PnC. 
For each measurement group, a single spectrum is then produced by taking the maximal value of the displacement noise at each frequency, giving the full spectrum of motion shown in Fig. \ref{fig:gen1_memb}c.
The simulated and experimental spectra are only in qualitative agreement; we attribute the discrepancies to microscopic differences in the geometry between the simulated model and the realized sample.
Such discrepancies however do not significantly affect the conclusions of this work.

The spectrum near DM A is characterized by two features which are absent for DMs B and C: first, a series of VEMs are measured at frequencies near that of mode A; second, we observe a splitting of the DM A into several peaks.
Such a concurrence is indicative of strong coupling between VEMs and DM A, a feature which has not been reported in other works\cite{Tsaturyan2017,Reetz2019}.
Note that we place particular emphasis on DM A because it is primarily studied in current applications of PnC membranes\cite{Reetz2019,Catalini2020}. 

\begin{figure}[bt]
\includegraphics[width=8.5cm]{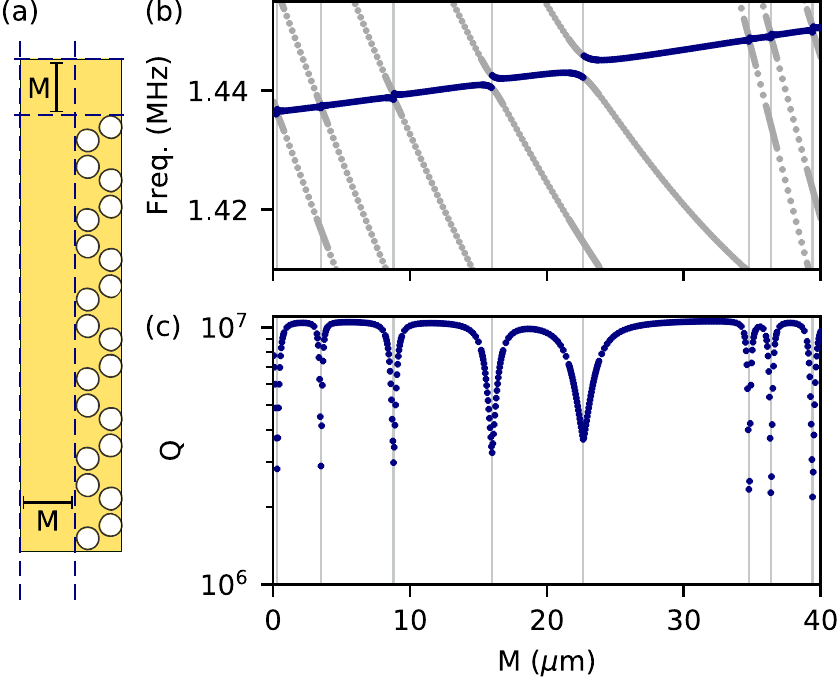}
\caption{\label{fig:simus}(a) Definition of the margin $M$ at the border of the membrane.
(b) and (c) Mechanical characteristics of DM A (blue) and nearby VEMs (light gray), as a function of the margin. 
Avoided crossings are marked by vertical lines.}
\end{figure}

To verify the strong coupling hypothesis, we first induce hybridization between the VEMs and DM A by precisely matching their frequencies in a numerical model with COMSOL.
The frequencies of the VEMs are found to vary strongly with a geometrical parameter of the membrane: the margin $M$, defined as the distance between any outermost hole of the PnC and the nearest border of the membrane -- see Fig. \ref{fig:simus}a.
We sweep this parameter such that several VEMs successively cross the frequency of DM A.
For each value of $M$, we extract the eigenmode frequencies (Fig. \ref{fig:simus}b) and their quality factor $Q$ (Fig. \ref{fig:simus}c).
The latter value is computed by using the definition $Q = Q_\mathsf{intr}\times U_\mathsf{elong}/U_\mathsf{bend}$.
The value of $Q_\mathsf{intr}$, which depends on the membrane thickness for sub-micron membranes where surface loss dominates, is taken to be $Q_\mathsf{intr}=6\times 10^3$ for $h=100$ nm, based on empirical measurements\cite{Villanueva2014}. 
$U_\mathsf{bend}$ is evaluated using Eq. (\ref{eqn:ubend}), whereas $U_\mathsf{elong}$ is computed with the formula $U_\mathsf{elong} = \rho h \Omega_\mathsf{m}^2\int_A dxdy \hspace{1mm} u_\sigma^2(x, y)/2$, where $\rho=3200$ kg.m$^{-3}$ is the density of SiN, and $\Omega_\mathsf{m}$ is the angular frequency.
This expression yields identical results to Eq. (\ref{eqn:uelong}) for high $\sigma$, but is more efficient for computation.
To further reduce the computation time, the system has been assumed symmetrical.
As a result, fewer VEMs appear compared to a more realistic model, but the general behavior of the system is maintained.

When the frequency of an edge mode approaches that of the DM, an avoided crossing can be observed, from which we estimate that the coupling rate between a VEM and the DM is of the order of a few kHz, a value consistent with the coupling rate recently reported in dimer defect membranes\cite{Catalini2020}.
The quality factor $Q$, shown in Fig. \ref{fig:simus}c, follows the behavior expected from a system of two coupled harmonic oscillators: the avoided crossings go along with a dramatic reduction in $Q$ by almost one order of magnitude.

To maximize the DM $Q$, one could in principle fine-tune $M$ to a specific value where the frequency difference between the DM an all VEMs is well above the coupling rate.
This is a challenging and non-reproducible approach because of the high accuracy in $M$ that is required, which exceeds the possibilities of the currently employed fabrication method.
Furthermore, any off-centering of the PnC pattern with respect to the membrane window will multiply the number of EMs as the symmetry is broken.
Another option is to reduce the coupling between VEMs and DMs by increasing the size of the membrane, and, correspondingly, of the PnC.
However, since the spectral density of EMs is proportional to the membrane side length and the coupling never strictly cancels, the risk of hybridization scales poorly with the PnC size. 

\begin{figure*}[bt]
\includegraphics[width=17cm]{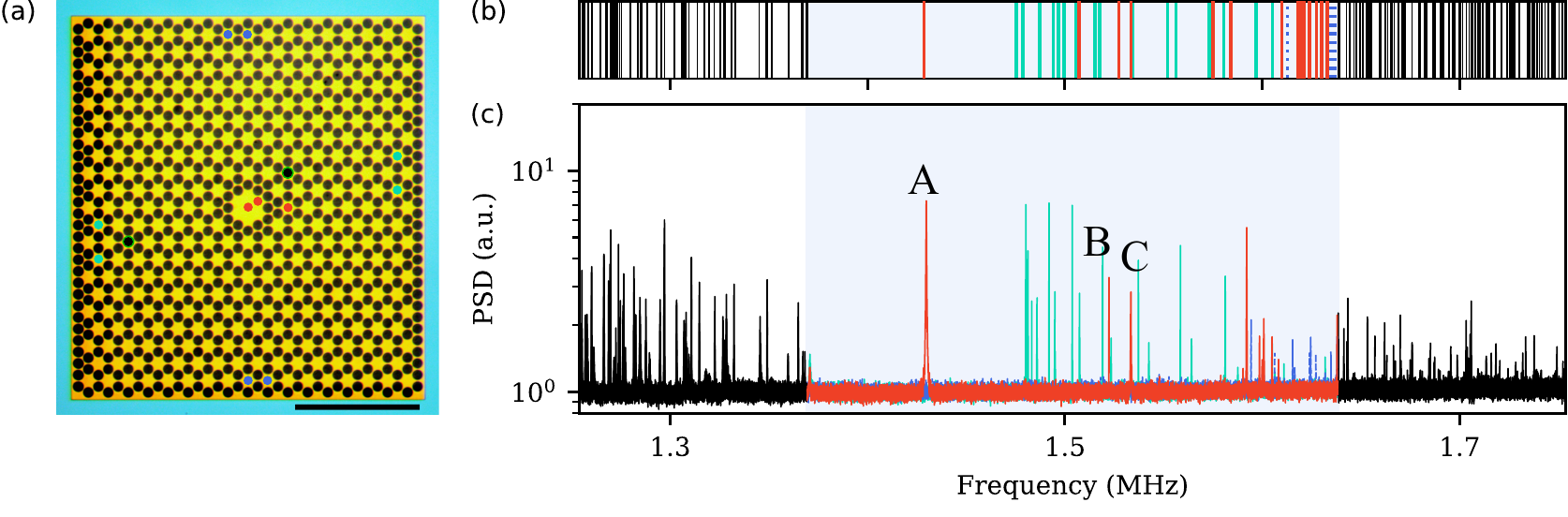}
\caption{\label{fig:gen2_memb}(a) Optical micrograph of a second generation PnC membrane. 
The dots represent the points where the spectra shown in (b) are measured. 
(b) Simulated spectrum near the PnC bandgap. 
(c) PSD spectrum consisting overall of 4 spectra, measured at the vertical borders (cyan), the horizontal borders (blue), the defect (red), and within the PnC (black).
For (b) and (c), the bandgap is highlighted in light blue.
}
\end{figure*}

For this reason, we suggest a third method, robust to errors in microfabrication, which will ensure reproducibly high quality factors for DM A.
Its principle is to engineer the frequencies of the VEMs to be far from that of DM A, ensuring that even with micro-fabrication errors, they are always sufficiently separated.
This second generation of membranes, shown in Fig. \ref{fig:gen2_memb}a, has its vertical borders patterned with supplementary arrays of holes near the anti-nodes of the VEMs.
This results in a shift toward higher frequencies of the VEMs, as shown in the simulated spectrum in Fig. \ref{fig:gen2_memb}b.
As before, qualitative agreement is found with the measured spectrum, shown in Fig. \ref{fig:gen2_memb}c.

Because of fabrication imprecision, the effect of edge mode coupling is not expected to be systematically observable on every sample.
For instance, no splitting of DM B or C can be observed in Fig. \ref{fig:gen2_memb}b, despite their spectral proximity to VEMs.
Rather, we expect that a good PnC membrane design would be robust to fabrications errors in $M$, resulting in a high reproducibility of the DM $Q$.
To verify this, we measure the fluctuations in $Q$ of the DMs of 9 samples for each membrane generation, with a design value of $M=20$ $\mu$m. 
The results for the first generation membrane are shown in Fig. \ref{fig:qs_database}a: DM A has an average quality factor $Q_\mathsf{A,1}$ of $3.9 \times 10^6$, with a standard deviation of $2.2 \times 10^6$. 
Equivalent values are $Q_\mathsf{B,1}=10.0 \times 10^6$ and $1.0 \times 10^6$ for DM B, and $Q_\mathsf{C,1}=9.6 \times 10^6$ and $1.0\times 10^6$ for DM C.
Furthermore, we check that the membranes with low $Q_\mathsf{A,1}$ indeed present mode hybridization, by imaging the mode profile.
In Figs. \ref{fig:qs_database}c and d, the profiles of two low-quality modes near the theoretical resonance frequency of DM A are shown: both of them present significant displacement in the central region and near the (left or right) vertical borders.
We attribute this breaking of left-right symmetry to a slight off-centering of the PnC with respect to the membrane window.

The results for the second generation membrane are shown in Fig. \ref{fig:qs_database}e: DM A has an average quality factor $Q_{A,2}$ of $7.2 \times 10^6$, with values spread over $2.0 \times 10^6$. Equivalent values are  $Q_{B,2} = 6.4  \times 10^6$ and $4.6\times10^6$ for DM B, and $Q_{C,2}=3.2 \times 10^6$ and $2.6 \times 10^6$ for DM C.
The quality factors in the absence of EM coupling (namely $Q_{A,2}$, $Q_{B,1}$, and $Q_{C, 1}$) agree to within 30\% with the optimal value predicted by our numerical simulations of $1.1 \times 10^7$.
These second generation membranes thus reproducibly attain the ideal dissipation dilution predicted by Eq. (\ref{eqn:q_qintr}) for DM A, which was the goal of our proposal, although this has the collateral effect of degrading DMs B and C.

The scaling of $Q$ with the thickness of the membrane emphasizes the importance of edge mode engineering: DMs follow a scaling described by Eq. (\ref{eqn:q_qintr}), while EMs, limited by anchor losses, follow the less favorable scaling of Eq. (\ref{eqn:qhc_qintr}). 
In our work, the aspect ratio $l/h\sim10^4$ leads to a tenfold reduction of the DM quality factor when hybridization with an EM occurs.
However, in experiments with more extreme aspect ratios up to $l/h\sim 10^5$ (with $h$ down to 12 nm)\cite{Rossi2018,Chen2019,Fedorov2020a,Galinskiy2020}, such events are expected to lead to a 100-fold reduction of the quality factor.

To summarize, we have presented our observation of a class of localized defect states of PnC membranes arising from finite size effects: edge modes.
We have demonstrated that they can hybridize with the high-$Q$ DMs, which degrades the performance of the membrane and reduces the reproducibility of the quality factors.
To address this issue, edge mode engineering has been used to ensure that the frequencies of the edges modes are maintained far from those of the fundamental defect mode, such that hybridization cannot occur due to fabrication error.
The method described here, by enabling the reliable fabrication of ultracoherent softly-clamped mechanical resonators, paves the way towards the wide-scale adoption of these devices in various scientific and industrial applications, such as magnetic resonant force microscopy \cite{Rugar2004,Fischer2019}, hybrid quantum networks\cite{Higginbotham2018,Midolo2018}, or cavity optomechanics\cite{Aspelmeyer2014}.

\begin{acknowledgments}
The authors would like to thank C. Chardin for early developments.
E.I. acknowledges support from the European Union’s Horizon 2020 Programme for Research and Innovation under grant agreement No. 722923 (Marie Curie ETN - OMT).
This work was supported by a starting grant from ''Agence Nationale pour la Recherche" QuNaT (ANR-14-CE26-0002).
\end{acknowledgments}

\section*{Data Availability statement}
The data that support the findings of this study are openly available on Zenodo at \newline http://doi.org/10.5281/zenodo.4055086, Ref.\cite{edgemodezenodo}.



\end{document}